*The Mnemosyne Number and the Rheology of Remembrance*

Safa Jamali[1,2] and Gareth H. McKinley[2]

[1]Department of Mechanical and Industrial Engineering,

Northeastern University, Boston, MA02115

[2]Hatsopoulos Microfluids Laboratory, Department of Mechanical Engineering,

Massachusetts Institute of Technology, Cambridge, MA02139

**Abstract**

The concept of a Deborah number is widely used in study of viscoelastic materials and to represent the ratio of a material relaxation time to the timescale of observation, and to demarcate transitions between predominantly viscous or elastic material responses. However, this construct does not help quantify the importance of long transients and non-monotonic stress jumps that are often observed in more complex time-varying systems. Many of these non-intuitive effects are lumped collectively under the term thixotropy; however, no proper nouns are associated with the key phenomena observed in such materials. Thixotropy arises from the ability of a complex structured fluid to remember its prior deformation history, so it is natural to name the dimensionless group representing such behavior with respect to the ability to remember. In Greek mythology, Mnemosyne was mother of the nine Muses and the goddess of memory. We thus propose the definition of a Mnemosyne number as the dimensionless product of the thixotropic time scale and the imposed rate of deformation. The Mnemosyne number is thus a measure of the flow strength compared to the thixotropic timescale. Since long transients responses are endemic to thixotropic materials, one also needs to consider the duration of flow. The relevant dimensionless measure of this duration can be represented in terms of a mutation number which compares the timescale of experiment/observation to the thixotropic timescale. Collating the mutation number and the Mnemosyne number, we construct a general two-dimensional map of thixotropic behavior, and quantify these ideas using canonical thixotropic models.

**Introduction**

In 1964, Marcus Reiner [1] during an after-dinner talk at the 4[th] International Congress on Rheology introduced the core rheological concept of the Deborah number (*De*) as a ratio of timescales; specifically the relaxation time of the material of interest as compared to the timescale of observation. The concept became central to the rheology community very quickly, and decades later, we commonly compare and contrast different timescales associated with rheological features of different linear viscoelastic material systems through this ratio, as well as other dimensionless groupings that quantify the additional (nonlinear) responses of more complex materials. For instance, in polymeric systems and processing flows, the Weissenberg number (*Wi*) provides an indication of how important non-linear rheological effects such as normal stress differences are in the material response of a complex fluid [2], while the Reynolds number (*Re*), familiar to many from fluid mechanics, provides a relative measure of fluid inertia to viscous stresses. Both of these latter parameters depend on the imposed flow strength; however, their ratio *Wi/Re* (commonly referred to as the elasticity number, *El*) is independent of the flow rate and provides a measure of the relative importance of viscoelastic stress relaxation to the viscous diffusion time. When considering hard-sphere suspensions under flow, one can compactly represent the rheological behavior with respect to the ratio between the rates of advection to the rate of diffusion, resulting in the Peclét number (*Pe*) [3]. In more complex attractive particulate systems, the Mason number (*Mn*), formed from the ratio of viscous shearing forces to attractive interparticle forces provides an effective dimensionless group to represent the quasi-steady state rheology of colloidal gels. Once again, the ratio of these two parameters, commonly denoted $\lambda = Pe/Mn$, is dimensionless and independent of flow strength, and provides a relative measure of how strong interparticle attractions are compared to the randomizing forces of Brownian motion. It is interesting to note that in both of these examples no proper names are associated with the [dimensionless] ratios of two [eponymous] dimensionless product groups.

Whilst these dimensionless groups have been extremely useful and important in categorizing the different regimes of material response observed in complex fluids and in confronting rheometric data with the predictions of appropriate constitutive models, they cannot help us in quantifying or describing the long transients and non-monotonic stress jumps observed in other more-complex time-varying systems. Many of these confusing and non-intuitive effects are lumped collectively under the term *Thixotropy* and the antonym *Rheopexy (or anti-thixotropy)*. This vast field has been reviewed extensively and authoritatively (see for example [4-11]); however, no proper nouns are commonly associated with key phenomena or underlying physical processes observed in such materials. This is perhaps due to conflation with a plethora of other rheological phenomena, such as dilatancy, shear-thinning, and viscoelasticity with thixotropy, which muddied the development of the field [12].

The term thixotropy was originally coined in 1927 by Peterfi [13], combining the Greek words *thixis* (touch) and *trepo* (turning), to describe the ability of living cells to re-gain their original solid-like state during a period of rest, after first being driven into a liquid-like state by agitation (a process we would now more appropriately refer to as *rheological aging*; see for example the clear discussion in [8]). Virtually all other early accounts of thixotropic behavior have to do with the sol-gel transition, and the ability of particulate gels to solidify (due to interparticle interactions) after becoming completely liquefied under flow [14-16]. In Larson's review [7] of thixotropic constitutive models, a clear definition and a distinction is presented for thixotropy vs. non-linear viscoelasticity: *"a time-dependent viscous response to the history of the strain rate, with fading memory of that history"*. Through the careful design of a suitable rheometric protocol (which we discuss at more length later), Divoux and coworkers [17] showed that this fading response can, in fact, be characterized by a thixotropic timescale (which they identify as $\theta$ but henceforth we shall denote $\tau_{thix}$ to be more consistent with SOR nomenclature), on which the material shows the most pronounced sensitivity to the history of deformation rate. Just as real viscoelastic fluids typically exhibit a spectrum of relaxation times, a typical thixotropic material may also show a range of thixotropic timescales [18]; however here, for simplicity, we assume that this distribution can be adequately and compactly represented by an appropriate moment or average of the underlying thixotropic time-scale spectrum.

There has been a re-awakening and growth of interest in thixotropy over the past few years, most probably because of its ubiquity in many real commercial systems. A concise but comprehensive summary of recent developments in the modeling of such systems has been given by Varchanis et al. [19]. Ewoldt and McKinley [20] have recently discussed three-dimensional phase map representations for thixotropic elasto-visco-elastic (TEVP) materials, in which plastic and viscoelastic behavior can be simultaneously observed with thixotropy; however, the lack of a clear and definitive nomenclature for the dimensionless groups to be used for parameterizing the magnitude of each effect results in ungainly terms such as the *thixoviscous number* and *thixoplastic number*. The goal of this short article is to propose suitable names for these dimensionless groups that can then be used by the rheology community to represent and discuss different aspects of thixotropic behavior in a clear and unambiguous manner.

## II. Mnemosyne: The Goddess of Memory

Astutely, when proposing the concept of the Deborah number to quantify phenomena being observed and reported in a dynamically-evolving field, Reiner avoided association with living rheologists, and we take the same approach here. We recognize that every thixotropic (or anti-thixotropic) effect observed in a complex fluid is owed to its ability to remember the history of its previous deformation. Hence it is natural to name the dimensionless group representing such behavior with respect to the ability to remember. In Greek mythology, *Mnemosyne* was one of the *Titans*, and the goddess of memory and remembrance. She presided over a river (or a spring) that flowed in parallel to the river of *Lethe*, the embodiment of forgetfulness. The dead, before reincarnation, drank water from the river Lethe to forget their past, in contrast to the idea of drinking from the Mnemosyne for novices in the Orphico-Pythagorean brotherhood [21]. In contemporary language we use the term *mnemonic* to refer to an artifact or device to help us remember a key concept or result. We thus propose to define the *Mnemosyne number* as the dimensionless product of the appropriately-defined thixotropic time scale of a material and the imposed rate of deformation: $My = \tau_{thix}\dot{\gamma}$. We propose the abbreviation *My* to avoid any confusion with the Mason number (*Mn*) that is already used in studies of colloidal gels. In any given kinematic situation, a high value of *My* indicates the potential for pronounced thixotropic memory of previous flow conditions and thus we expect the Mnemosyne number to capture (at least in part) a suitable measure of the extent of thixotropy in a systems.

The value of the Mnemosyne number allows us to clearly distinguish thixotropic phenomena from other rheological responses; and any rheometric test or processing flow of interest may correspond to vanishingly small values of *Re*, *De* and/or *Wi*, but at a finite (and perhaps large) value of *My*. For instance in shearing of carbon black or clay suspensions, non-linear viscoelastic effects (such as normal stress differences) are typically much smaller in magnitude and less important than thixotropic transients, so $Wi << My$. The ratio $Wi/My = \tau_{el}/\tau_{thix}$ has been previously referred to as a *thixoelastic parameter* by Ewoldt and McKinley [20] and helps distinguish whether nonlinear elastic effects are more important ($Wi/My >> 1$) or thixotropic/aging effects dominate ($Wi/My << 1$). Experimental protocols for distinguishing such effects have recently been considered by Agarwal et al. [11]. We do not seek to rename this quotient because, like the quantities *Wi/Re* and *Pe/Mn* discussed above, it is a deformation-rate–independent quantity formed from the ratio of two extant, and already named, dimensionless parameters.

The Mnemosyne number, as defined here, is a measure of flow *strength* compared to the thixotropic timescale. However, because of the long transient responses endemic to thixotropic materials, one also needs to consider the duration of any flow protocol in order to determine how large the magnitude of the actual thixotropic rheological response is. In dealing with many thixotropic materials, the experimental protocol commonly involves a pre-shearing state at large deformation rates for a carefully-prescribed period of time in order to erase the material's memory of its previous flow history (e.g., the initial trauma of loading it into a rheometer). It is appropriate thus to refer to this pre-shearing process as "*letherizing*" the material sample, as the memory of its previous life(s) is forgotten. For instance, in order to fully letherize a thixotropic sample, *strong flows of long duration* are required. The relevant duration of a flow or rheometric test protocol (which we denote for clarity as $t_{exp}$) for a time-evolving material has been considered quite generally by Mours and Winter [22], and they argue that this can be represented in terms of a *mutation number*: $Mu = t_{exp}/\tau_{thix}$, which compares the duration of the experiment/observation to the timescale characterizing the rate of change of the material (here the thixotropic time scale $\tau_{thix}$. For example, in order to ensure that a viscoelastic fluid doesn't change its properties significantly or evolve during a typically oscillatory test (for which the duration is $t_{exp} \approx 2\pi/\omega$) we require *Mu* << 1. This constraint has prompted the development of new fast oscillatory shear test protocols such as the Optimized Windowed Chirp [23].

### *III. Mutation-Memory Maps*

Combining these considerations regarding both the strength and the length of a shearing protocol, it becomes clear that the Mnemosyne number and the mutation number can be used to construct a general two-dimensional map of thixotropic behavior as shown schematically in Figure 1. First, we note that during a specific time-dependent flow protocol (e.g. start-up of steady shear flow), the product of these two dimensionless groups gives a measure of the total accumulated strain: $My.Mu = (\tau_{thix}.\dot{\gamma})(t_{exp}/\tau_{thix}) = \gamma$, as indicated by the broken hyperbolic lines. Low strains correspond to the lower left and large strains to the upper right of this operating space. We now consider the physics captured in different quadrants of this thixotropy map. The viscoelastic response of a TEVP material can be reliably measured using weak flows of (relatively) short duration located in the lower left of the map ($Mu \ll 1$, $My \ll 1$). Here thixotropic changes to the material as well as the total shearing deformation imposed are rather small and the sample does not mutate during the experimental test. In contrast, very strong and long flows in the upper right ($Mu \gg 1$, $My \gg 1$). are used to letherize the material into a fluid-like behavior and eradicate all memory of the previous states of the material.

At the heart of this map at intermediate total strains lie regions with pronounced thixotropic and hysteretic effects. This is the region where long-duration – and frequently-confounding – transient dynamical responses, such as non-monotonic stress evolution and/or transient shear banding are observed. We explore this regime in greater detail below. To the far right of this figure at small $My$, and large $Mu$, (corresponding to weak thixotropic effects and longer observation times), one can recover the quasi-steady flow curve of a material, provided the accumulated strain is also large enough that the system evolves (or mutates) towards its steady flowing state. Conversely, very large values of $My$ and very small $Mu$ (corresponding to the upper left of Figure 1) correspond to the process of quenching a time-dependent thixotropic material into a non-equilibrium (often glassy or non-equilibrium gel) microstructure.

Finally, for completeness, we note that other time-dependent effects such as "rheological aging" of glassy materials are possible even under rest (no imposed flow) conditions. These effects are clearly distinguished from thixotropy by Wei and Larson (see Section II.B of [8]). Such phenomena (corresponding to $\dot{\gamma} = 0$) evidently cannot be uniquely represented as coordinate points on a two-dimensional map of the form of Figure 1 and would necessitate a more complex (but straightforward) three-dimensional representation, which we don't pursue further here. The dynamic rheological process commonly known as "rejuvenation" would correspond to projection of a material's coordinates from high (i.e. "old") values of this third (as yet unnamed) material age axis back to low values characteristic of a rheologically-young material.

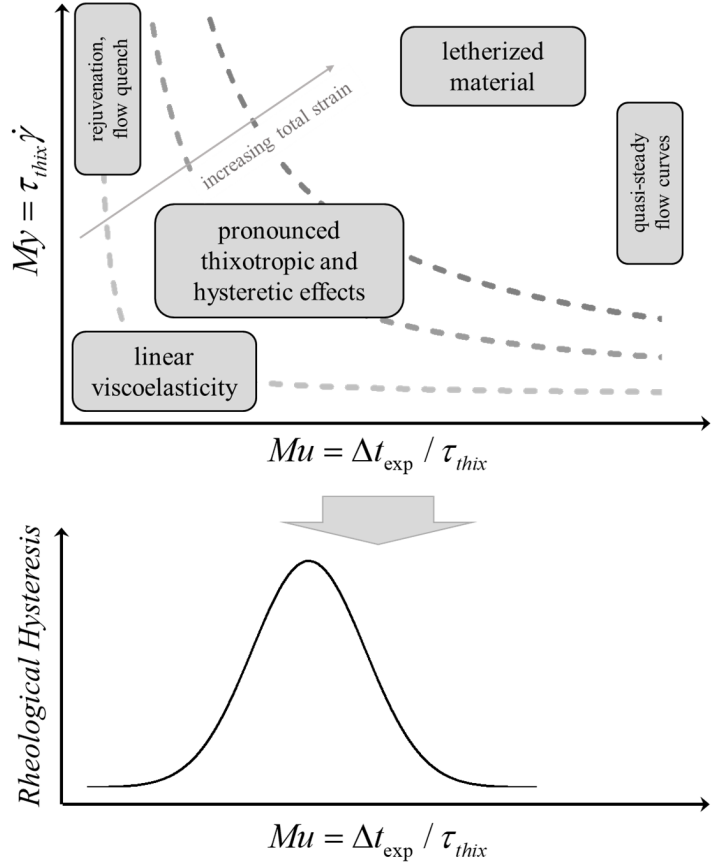

Figure 1. A proposed phase map for representing, locating and understanding common rheological phenomena associated with rheological characterization of thixotropic materials in terms of the Mnemosyne number, *My* and the mutation number, *Mu*. The lower figure shows a generic measure of the non-monotonic evolution rheological hysteresis (see text for suggested definition) as a function of the experimental duration (or mutation number) of the experiment.

As we have depicted schematically in the lower part of Figure 1, quantitative measures of the *rheological hysteresis* that arises during shearing of a thixotropic material are expected to be a function of the timescale of the experimental test protocol (or, equivalently, the mutation number *Mu*) and will be most pronounced in intermediate strain/time regimes. To generate quantitative and reproducible thixotropic data to map out this non-monotonic hysteresis curve it is essential to first develop a robust rheometric test protocol. Early experimental iterations of specialized instruments such as the *Thixotrometer* of Pryce-Jones [24] often used "thixotropic loops" starting from rest conditions and first increasing (and then decreasing) the imposed shear rate (or shear stress) and are discussed in detail by Bauer & Collins [12]. However, for many structurally-sensitive commercial systems, as well as model colloidal "soft glassy" materials, the initial state of the sample is a strong function of the loading/preparation history, as well as the waiting time $t_w$ before the experiment is performed. This can make it hard to achieve repeatable and device-independent data. Mewis and coworkers [5] have argued convincingly for the adoption of 'step rate' tests (in which the shear rate on the material is jumped rapidly from a low value to high, or vice versa). Performing a series of such step-rate tests provides a rich dataset that indeed maps out the functional describing the entire thixotropic material response envelope, but can be extremely time-consuming to generate and analyze. In a very recent paper, Choi *et al.* [25] used a series of stress jump experiments and mapped thixotropic behavior to distinguish between elastic and viscous contributions to the total fluid stress.

Recently, Divoux and co-workers [17, 26] have described a somewhat simpler but robust ramp-down/ramp-up flow protocol that provides the data density required to probe the spectrum of thixotropic responses in a material in a more efficacious way. One begins in the upper right of Figure 1 at an initially large shear rate, $\dot{\gamma}_i$, corresponding to a *fully letherized* material (thus generating a repeatable and history-independent initial configuration) and then imposes a series of different shear rates down to a final minimum shear rate, $\dot{\gamma}_f$ before reversing the process and returning back to the initial (large) shear rate. These down/up ramps may consist of a series of $n$ discrete steps, each of length $\delta t$ (as utilized in [17] so that the total elapsed time is $t_{\exp} = n\delta t$) or, perhaps more conveniently, a single continuously-varying ramp down at a specified rate, $r$. This continuous ramp could also be chosen to be linear, or a power-law or exponential in character. In recent DPD computations with an attractive colloidal system [27, 28] we have shown that although the precise numerical values of the computed hysteresis measure will change for each ramp protocol, the qualitative features of this protocol are robust and quite generally of the nonmonotonic form sketched in Figure 1.

In Figure 2 we illustrate a number of different possible protocols, namely a step-wise (blue) and a continuous ramp down/up (red), a single step jump or "quench" protocol to a low final shear rate (orange curve) and finally a simple start-up of steady shear flow protocol (as shown by the green step function) at an arbitrary deformation rate. The top figure depicts the temporal variation in the applied deformation rate of these different protocols for a given (constant experimental duration $\Delta t_{exp}$) while the bottom figure shows the corresponding graph mapped onto the *My-Mu* diagram with direction indicated by an arrow. Different levels of shading/transparency increments show different (increasing) values of $\Delta t_{exp}$. Note that the initial departure point on the *My-Mu* diagram for each of these protocols is selected to always lie within the fully letherized state (so that all previous deformations have been forgotten). The stress jump and quench, step-wise and continuous ramp protocols for a given $\Delta t_{exp}$ are indistinguishable on this diagram. Thus for illustration purposes, different circular points indicating different values of the prescribed shear rate steps are shown on top of the continuous line for ramp down/up protocols, and a dashed line (with no intermediate points and just a final state at $\dot{\gamma}_{min}$) is presented for the stress jump/quench protocol. In contrast, the start-up of steady shear experiments (shown by green points) are represented by a series of horizontal points on the *My-Mu* diagram, as the applied shear rate (and thus the Mnemosyne number), is constant throughout, and the experimental time that enters the numerator of the mutation number increases monotonically with the time of shearing. It is important to note that the entire *My.Mu* space can still be explored using this protocol by performing a series of start-up experiments at different values of the imposed shear rate – corresponding to a series of horizontal sweeps through this *My.Mu* parameter space.

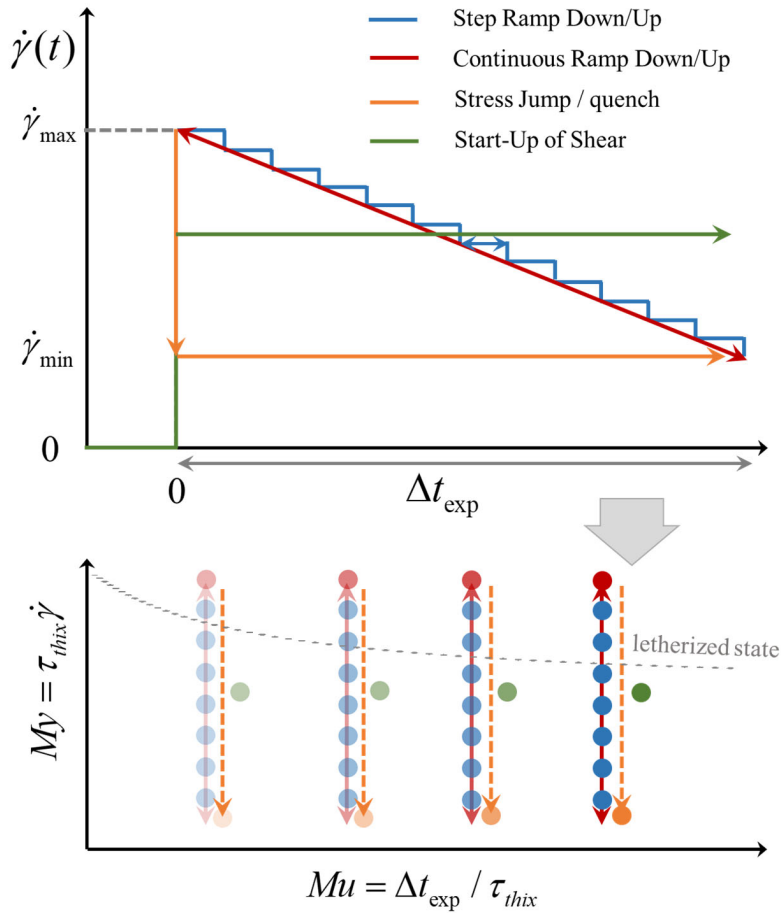

Figure 2. The time-varying shear rate as a function of time for a given $\Delta t_{exp}$ for different flow protocols (top), and corresponding values mapped onto the proposed *My-Mu* diagram. Different color transparencies indicate different values of $\Delta t_{exp}$.

For didactic simplicity we only consider in detail here a linearly-varying shear rate $\dot{\gamma}(t) = \dot{\gamma}_i - rt$ performed at a (user-selectable) rate $r$. The total time of the down-ramp experiment is then $t_{exp} = (\dot{\gamma}_i - \dot{\gamma}_f)/r$ and consequently the kinematic history of the down-ramp test can also be written as $\dot{\gamma}(t) = \dot{\gamma}_i - (\dot{\gamma}_i - \dot{\gamma}_f)\frac{t}{t_{exp}}$. A typical protocol may select $\dot{\gamma}_i = 100 s^{-1}$ and $\dot{\gamma}_f = 1 s^{-1}$, followed by an up-ramp with the initial and final shear rate values interchanged. As proposed by Divoux et al. [17] and adapted by [26-28] an appropriate measure of the rheological hysteresis arising from thixotropy is then the difference in areas under the apparent (non-equilibrium) flow curve of $\sigma(\dot{\gamma})$ measured on the up/down trajectories:

$$A_{\Delta\sigma} = -\int_{\dot{\gamma}_f}^{\dot{\gamma}_i} \sigma^{(down)} \, d\log\dot{\gamma} + \int_{\dot{\gamma}_f}^{\dot{\gamma}_i} \sigma^{(up)} \, d\log\dot{\gamma} \equiv \int_{\dot{\gamma}_f}^{\dot{\gamma}_i} \Delta\sigma(\dot{\gamma}) d\log\dot{\gamma} \qquad (1)$$

Where $\Delta\sigma(\dot{\gamma}) = \sigma^{(up)} - \sigma^{(down)}$, and can in principle be positive or negative (for an anti-thixotropic material). To avoid ambiguity and focus on the magnitude of the hysteretic effects, we follow [17], and use the absolute magnitude of the stress difference, $|\Delta\sigma(\dot{\gamma})|$, in up/down ramps. In this formalism the shear rate is logarithmically spaced so that an equal weight is given to low and high shear rates [17]. Nonetheless, the hysteresis area as defined in eq.(1) is not dimensionless and we therefore normalize the hysteresis area by the area under the ramp-down flow curve (also schematically shown in the bottom right sketch of Figure 3), resulting in the following definition of a dimensionless hysteresis measure that captures the extent of thixotropy in a material

$$\tilde{A}_\sigma = \frac{\int_{\dot{\gamma}_f}^{\dot{\gamma}_i} |\Delta\sigma(\dot{\gamma})| d\log\dot{\gamma}}{\left|\int_{\dot{\gamma}_i}^{\dot{\gamma}_f} \sigma^{(down)} \, d\log\dot{\gamma}\right|} \qquad (2)$$

One could alternatively non-dimensionalize the hysteresis area defined in eq.(1) by scaling the shear stress values by a yield stress $\sigma_y$ and normalizing the shear rate values by $\dot{\gamma}_f$; however, not all thixotropic fluids exhibit a yield stress, and even for the ones that do, from an experimental point of view the precise value of the yield stress is not necessarily always clear.

By imposing a series of different up/down ramp rates $r_k$ (for $k = 1,2...N$) as shown schematically in Figure 3 we can then systematically map out the (typically) non-monotonic evolution of the hysteresis area in the apparent flow curve $\sigma(\dot{\gamma}(t;r))$ and identify the characteristic thixotropic time scale $\tau_{thix}$ of the test material. For very long duration experiments (with large values of $t_{exp}$) the quasi-static nature of the experiment means that the equilibrium flow curve is obtained with little/no hysteresis. However, as the ramp rate down/up is increased, the experimental duration becomes progressively shorter and rheological hysteresis becomes significant, if the sample is thixotropic.

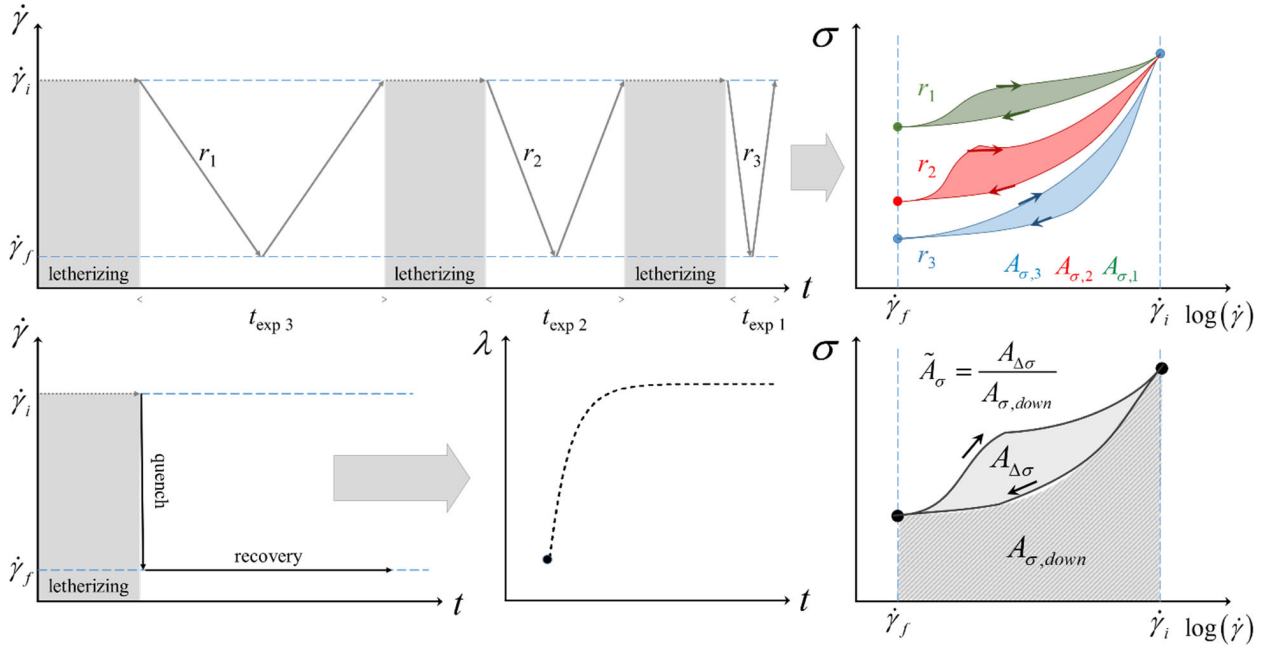

Figure 3. An example of: (top) a repeated up/down ramp protocol, and (bottom) a stress (or shear rate) jump experiment, for probing rheological hysteresis and identifying the thixotropic time scale of a material. The quench experiment provides a convenient protocol for monitoring the recovery of the material (as represented here schematically by the evolution in a scalar structural parameter $\lambda(\dot{\gamma},t)$). Between each ramp $r_k$ ($k$ = 1, 2, 3...) a letherizing period at high shear rate ($\dot{\gamma}_i$) rejuvenates the material. The high value of the Mnemosyne number $My_i$ ensures elimination of any shear-induced structure (SIS) formation, and monitoring the time-invariance of $\sigma(\dot{\gamma}_i)$ also allows us to check that no long-term irreversible effects such as sample drying, or progressive shear-induced sample degradation are in play. On the right we have sketched the corresponding apparent flow curves $\sigma(\dot{\gamma};r)$ that might be expected for a thixotropic material with a yield stress (for a recent example see [29]), but the existence of a yield stress is not essential.

This time-varying deformation protocol is thus another example of *mechanical spectroscopy,* in the same way that (for a non-mutating, viscoelastic material) imposing a sequence of different oscillatory deformation frequencies (either as a set of discrete frequencies with $\tau_{exp} \approx 2\pi/\omega$ or in a single, time-varying "chirp" waveform) probes different well-defined states – such as the terminal regime (at low frequencies), the rubbery plateau (at high frequencies) – as well as enabling determination of the full relaxation spectrum of the material.

Very rapid ramp rates down from $\dot{\gamma}_i$ are akin to mechanically "quenching" the microstructure of the material in its fully letherized (shear-rejuvenated) state. A useful analogy here is the rapid thermal quenching of molten metals in order to form metallic glass states with markedly different mechanical properties, as compared to slowly-cooled equilibrium microstructures that feature polycrystalline domains. Recent simultaneous measurements of microstructure or conductivity and dynamic modulus in carbon black pastes being developed for battery slurries have also illustrated the variety of non-equilibrium states that can be developed in thixotropic microstructured materials [30]. Of course, we must bear in mind that some complex fluids may have such short thixotropic time scales that it is not possible (within the limitations of current rheometers) to quench the structure fast enough to explore the full non-monotonic curve of $A_\sigma(r)$. This seems to be the case for carefully-prepared Carbopol samples which have a thixotropic time scale of less than one second [17] and thus always reside towards the far right of Figure 1. Similarly we may never be able to shear a material strongly enough to fully letherize it and completely eliminate memory of its previous lives (see the discussion in Section II.D of [8]).

## IV. Rheological Hysteresis in Common Models for Thixotropy

Finally, we note that this framework is also of use in understanding the response of constitutive equations written for thixotropic materials. Virtually all inelastic constitutive equations, beginning with the expression proposed by Goodeve [31] to enhanced variants such as those discussed by Coussot and Bonn [32], to detailed phenomenological viscoelastoplastic models such as the Isotropic-Kinematic Hardening (IKH) model [33-36], resort to a description of the evolution in one (or more) scalar thixotropy parameter(s), $\lambda(t;\dot{\gamma}(t'))$, which capture some appropriately-weighted and normalized measure of the evolving distribution in the microstructural states within the material. The evolution in this parameter then controls the bulk rheological response of the material. Early computational explorations of this form (including up/down hysteresis loops) can be found in the pioneering work of Frederickson [37].

In the simplest generic form, one can construct a time-evolving functional for this structural parameter that might be written generically as:

$$\frac{d\lambda}{dt} = \frac{1}{\tau_{thix}}(1-\lambda) - \beta\lambda\dot{\gamma} \qquad (3)$$

The first term on the right describes "creation" or rebuilding of microstructure, while the second is a shear-rate–dependent destruction term. The build-up rate is determined by the characteristic thixotropic time of the material, and $\beta$ is a (material-dependent) dimensionless parameter describing how effectively the microstructure is broken down under shear. A simple steady-state solution of such an equation [at a given deformation rate] yields: $\lambda = \frac{1}{1+\beta My}$, clearly illustrating the role of the Mnemosyne number in quantifying changes in the microstructural state and (ultimately) the bulk rheology of a thixotropic material under shear. A full time-dependent solution of eq. (3) for a time-varying up-down ramp $\dot{\gamma}(t;r)$ corresponds to a <u>vertical</u> trajectory through the state map in Figure 1; starting at the top (*i.e.,* at a high initial shear rate with $My_i \gg 1$) and moving first down (*i.e.,* a ramp down in shear rate to a final value $\dot{\gamma}_f$ with $My_f \ll 1$) and then back up (*i.e.* a ramp up in shear rate). The specific value of $Mu$ for this test is set by the specified ramp rate $r$. The degree of hysteresis measured (or computed) during this down/up ramp will of course depend on the ramp rate $r$ and the specific form of the thixotropic constitutive model under consideration (of which there are myriad; see, for example, the discussion/review of Varchanis et al [19], or [38]).

Here we have performed a simple numerical investigation of the flow protocol mentioned above, $\dot{\gamma}(t) = \dot{\gamma}_i - (\dot{\gamma}_i - \dot{\gamma}_f)\frac{t}{t_{exp}}$, with eq. (3) governing the time evolution of the thixotropy parameter, $\lambda(t;\dot{\gamma}(t))$. One then can select or construct different constitutive equations for the stress response of the fluid under such flow protocols. We present three simple limiting cases of appropriate thixotropic constitutive equations:

(i) An inelastic *thixo-viscous* (TV) *fluid* in which the shear stress response to an applied deformation rate is given by: $\sigma(t) = [\eta_s + \eta_p \lambda(t;\dot{\gamma}(t))]\dot{\gamma}(t)$. Here the total viscosity of the thixotropic material is written as a sum of contributions from an invariant background Newtonian solvent viscosity, $\eta_s$, and from the evolving structural viscosity, $\eta_p \lambda(t)$.

(ii) A *thixo-plastic* (TP) *fluid* with the constitutive equation given as: $\sigma(t) = \sigma_y \lambda(t; \dot{\gamma}(t)) + \eta_s \dot{\gamma}(t)$, for which the value of the yield stress, $\sigma_y$, depends on the level of structure in the material but the plastic shearing viscosity remains constant.

(iii) A *thixo-visco-plastic* (TVP) *fluid* (which might also be referred to as a Thixotropic Yield Stress Fluid or TYSF [29]) that takes a constitutive form combined from (i) and (ii) as: $\sigma(t) = \sigma_y \lambda(t; \dot{\gamma}(t)) + \left[\eta_s + \eta_p \lambda(t; \dot{\gamma}(t))\right]\dot{\gamma}(t)$. For a TVP fluid, the evolution of the microstructure characterizing the material directly changes the yield stress as well as the plastic viscosity of the fluid.

Of course there are a myriad of other functional forms, some of which have been developed to include the role of elasticity as well [33, 36, 39]; however, here, we focus on a simple class of inelastic models with no elastic response. Note that for all three formulations examined here, a material with no microstructure, $\lambda = 0$, corresponds to a simple Newtonian rheology. In the numerical calculations described below we fix the value of the yield stress at $\sigma_y = 20$ Pa, the Newtonian viscosity at $\eta_s = 1$ Pa.s and the structural viscosity at $\eta_p = 4$ Pa.s consistent with the carbon black gels considered by Helal *et al.* [30]. Finally for illustrative purposes we set $\tau_{thix} = 10$ s consistent with the results of Divoux *et al.* [17]. The transient solution for the evolution in the thixotropy parameter (eq. 3) is independent of the choice of stress constitutive equation and is presented in Fig. 4 for a prototypical choice of $\beta = 0.1$ and a very wide range of experimental durations varying from $10^{-2} - 10^{6}$ s. The solid lines represent the ramp-down flow from the initial shear rate of $\dot{\gamma}_i = 100$ s$^{-1}$ to the final shear rate of $\dot{\gamma}_f = 0.1$ s$^{-1}$ and the dashed lines represent the ramp-up protocol back to the fully-letherized or destructured state (often also refered to as "shear melting"). There are many features that are clearly evident in the thixotropic structural evolution curves shown in Fig. 4. Firstly, one can immediately identify hysteresis loops as the ramp up/down curves do not collapse onto each other for intermediate experimental durations. For very short flow durations, $Mu \approx 0$, the material does not have sufficient time to build up structure and thus $\lambda$ remains virtually null for the duration of the experiment, resulting in a negligible hysteresis area. On the other hand, for very long experimental durations, $Mu \gg 1$ and for each instantaneous shear rate the thixotropy parameter evolves to its quasi-steady state value. Thus the hysteresis area is minimal or non-existent for this regime as well. The parameter $\beta$ directly controls the effectiveness of shear flow in breaking up the structure and thus controls the quasi-steady state value of $\lambda$ to be $\lambda(t) \approx \dfrac{1}{1+\beta\tau_{thix}\dot{\gamma}(t)} = \dfrac{1}{1+\beta My(t)}$. For our choice of parameters this yields $\lambda_f \approx 0.9$ at the end of the ramp down phase (when $\beta My(t_f) = \beta\tau_{thix}\dot{\gamma}_f = 0.1$), consistent with the results shown in Fig. 3.

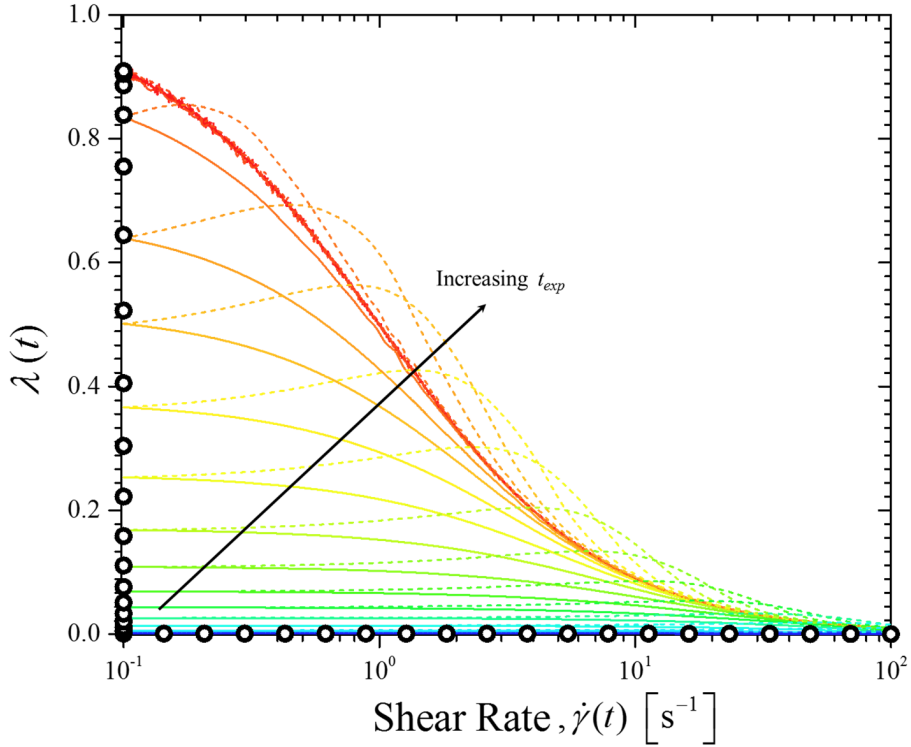

Figure 4. Numerical solution of Eq. (3) for time evolution of the thixotropy parameter, $\lambda(t,\dot{\gamma})$, under a linear ramp down/up flow protocol as shown schematically in Figure 3. The color increments from blue to red represent progressively longer total times of experiments, $t_{exp}$, corresponding to different mutation numbers ranging from $Mu = 10^{-3} - 10^5$. The solid lines present the initial ramp-down, and dashed lines represent the subsequent ramp-up flow protocol for the same experimental duration. An initial value of $\lambda = 0$ is assigned to the fully-letherized state at the initial high shear rate ($\dot{\gamma}_i$). The black hollow circles spaced along the bounding axes represent the trajectory of a sudden flow cessation experiment, in which the shear rate is reduced from $\dot{\gamma}_i$ to $\dot{\gamma}_f$ instantaneously (i.e. the flow is "quenched") and kept at the lowest value for the remainder of the experiment, while monitoring the subsequent recovery in the thixotropy parameter.

Knowing the temporal evolution in the thixotropy parameter $\lambda(My,t)$, one can subsequently compute the expected rheological hysteresis by looking at the evolution in the flow curves for the three different subclasses of thixotropic fluid model (TV, TP and TVP) summarized above. In Figure 5 we show the shear stress response of these different models for the same set of ramp-down/ramp-up protocols used for computing the thixotropy parameter presented in Fig. 4. The inset figures show the value of the hysteresis area, $A_\sigma(t_{\exp})$ as the total time of the experiment changes for three different values of $\beta = 0.01$, $0.1$, and $1$. Firstly, it is clear that all three models robustly predict non-monotonic hysteretic curves with clear maxima in their distribution of $A_\sigma(t_{\exp})$. This clearly demonstrates that the hysteresis observed in the thixotropy parameter itself is sufficient to lead to rheological hysteresis. For instance, one does not need to necessarily have a yield stress fluid to observe rheological hysteresis (and indeed some complex foodstuffs such as Marmite appear to be *thixoviscous* in nature [40]). However, a closer look shows that the precise form of the constitutive model can substantially change the quantitative characteristics of the $A_\sigma(t_{\exp})$ distributions.

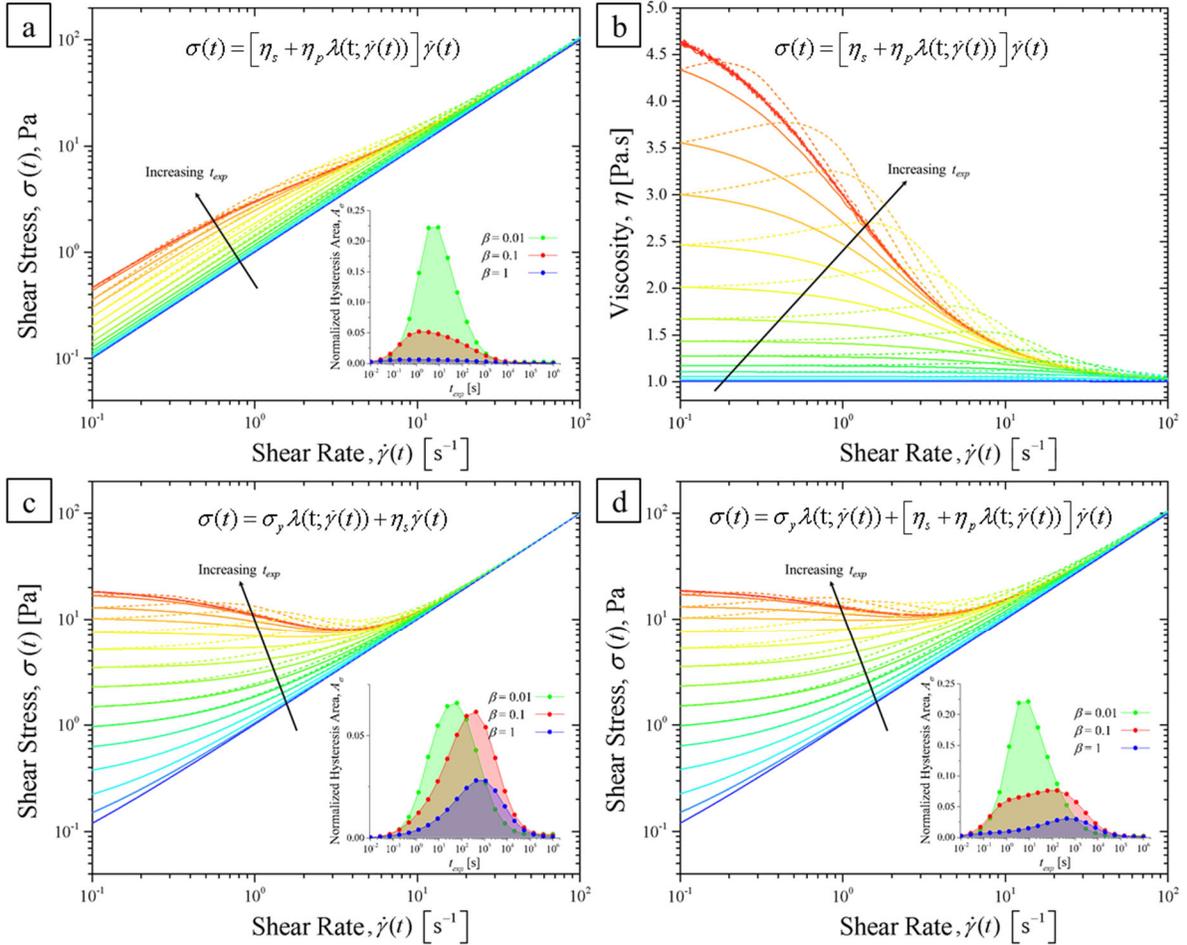

Figure 5. (a, c, d) Expected evolution in the measured shear stress vs. applied shear rate, under the ramp-down/ramp-up flow protocol shown schematically in the top row of Fig.3. Three different types of thixotropic fluid model are considered: (a, b) TV, (c) TP, and (d) TVP, with constitutive equations for each model shown in the figure, and the results are evaluated numerically using the time-dependent solution for the thixotropy parameter presented in Fig. 4. For clarity in (b) the viscosity of the purely thixoviscous (TV) fluid is also presented as a function of shear rate, resulting in identical hysteresis loops to those shown in fig. 3 (due to the linear dependency of the total viscosity on the evolving thixotropy parameter). The color increments from blue to red represent longer total times of experiments, $t_{exp}$, corresponding to different $Mu$ numbers in the range $Mu = 10^{-3} - 10^{5}$. The total hysteresis areas $A_{\sigma}(t_{exp})$ computed using the expression in eq. (2) are calculated for different experiment times, and different values of the parameter $\beta$, and are presented as inset figures for each constitutive model. The solid lines present the initial ramp-down flow, and dashed lines represent the ramp-up flow protocols for the same experimental durations.

For a given value of $\beta$ (here taken to be $\beta = 0.1$), we can compare and contrast the hysteresis areas measured from the flow curves for three different fluids above, as shown in Figure 6. First it should be noted that, for all three different fluid models considered, the value of the thixotropic timescale in these calculations (held constant here at $\tau_{thix.} = 10$ s) differs from the characteristic value of $t_{exp}$ at which the local maximum in the distribution curve of the hysteresis area is observed. This is to be expected because: (i) the value of $\beta$ clearly influences the location of the hysteresis maximum as evident in the inset figures shown in Fig. 5, (ii) for the TV and TVP fluid models the value of the plastic viscosity can change the exact location of the hysteresis maximum. We also note that in the original step-wise flow protocol described by Divoux *et al.* [17] (and later adapted by [26, 28]), the time scale presented on the ordinate axis is the time spent at each shear rate ($\delta t$), as opposed to the total time of the experiment ($t_{exp} = n\delta t$) in our proposed flow protocol.

Additionally, one should note that by defining the hysteresis area as given in eq. 2, the magnitude of the shear stress enters the expression, resulting in larger weighted contributions to the total hysteresis area when thixotropic effects are more pronounced at high shear rates.

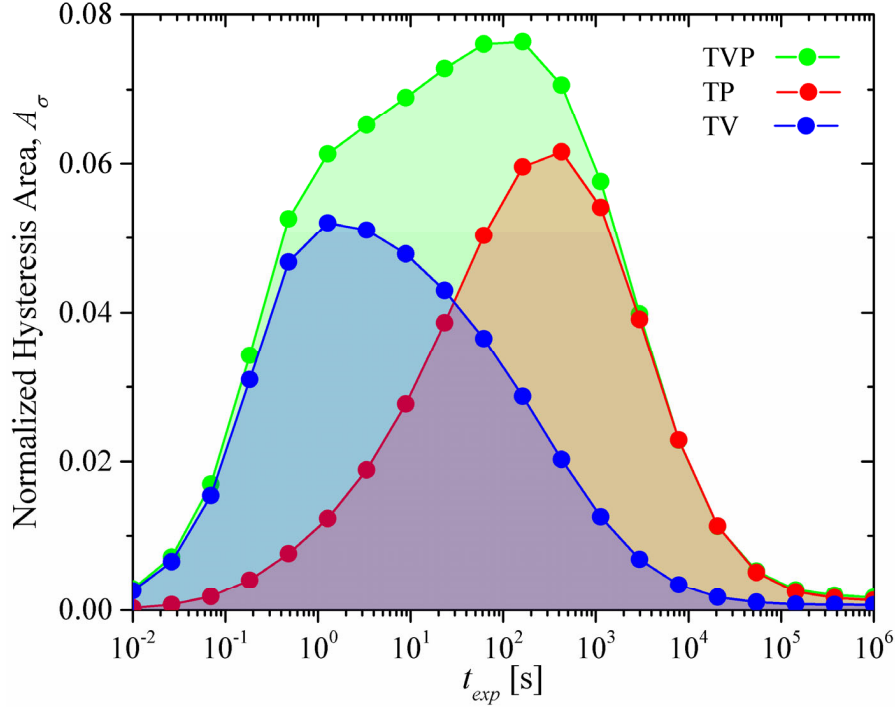

Figure 6. Total rheological hysteresis computed for the three different types of inelastic thixotropic fluid models, all with the same constant values of $\beta = 0.1$ and $\tau_{thix} = 10\,\text{s}$ in equation 2.

If instead computation of the hysteresis area was based (for example) on $\log(\sigma)$ as $A_\sigma = \int_{\dot{\gamma}_f}^{\dot{\gamma}_i} |\Delta \log \sigma(\dot{\gamma})| \, d\log \dot{\gamma}$ this would shift the location of the maxima in the distribution towards shorter times. For completeness we show such a plot in the Electronic Supplemental Material

A close inspection of the results for the TP and TV fluid models shows that although the choice of parameters can increase/decrease the total hysteresis area, and the precise value of the experimental timescale at which the maximal hysteresis observed, in general a rather simple distribution with a clear single maximum is measured for each model. This is in agreement with measurements in a range of different yield stress materials [17] and also with DPD computations [27]. In contrast, the results for the TVP fluid model show a more complex bimodal distribution with distinct contributions from both the viscous and the plastic contributions to the stress. Other more sophisticated models such as those proposed by Geri *et al.* [35], Larson [8, 18], and a population balance model proposed by Mwasame *et al.* [41], in which different microscopic physical processes (with different characteristic time scales) give rise to different contributions to the total stress will also predict a similar multi-modal response.

Because of this difficulty in directly recovering the precise numerical value of the thixotropic time scale (which in the three models considered here is of course known *a priori*) from hysteresis area measurements, we propose augmenting the ramp-down/ramp-up protocol with one simple final additional step; *i.e.* a sudden cessation or "quench" of the system from an initially-letherized state (corresponding to a long time of shearing at a high shear rate $\dot{\gamma}_i$ (i.e. $Mu \gg 1$ and $My \gg 1$) to the lowest possible final measurable shear rate ($\dot{\gamma}_f$), and directly monitoring the recovery in the shear stress as the microstructure rebuilds. We perform this stress jump experiment numerically with a very short ramp down protocol (corresponding to $t_{exp} = 0.1$ second) followed by 100 seconds of steady shear flow at a constant shear rate of $\dot{\gamma}_f = 0.1 s^{-1}$. Results of such a quench/recovery protocol are presented as the hollow black solid circles in Fig. 4 (holding the shear rate at a constant low (but non-zero) value results in evolution of the thixotropy parameter in a vertical fashion along the ordinate axis when results are plotted vs. the applied shear rate). The same results are presented against time in Figure 7 and show a smooth monotonic recovery in the structure and thus in the shear stress of the material as well. Equation 2 can also be solved analytically using an integrating factor when the shear history $\dot{\gamma}(t)$ is specified. For the simplest case, when a constant shear rate of strength $\dot{\gamma}_f$ is applied this results in the following expression:

$$\lambda(t;\dot{\gamma}_f) = \frac{1}{1+\beta\dot{\gamma}_f \tau_{thix}} - \left(\frac{1}{1+\beta\dot{\gamma}_f \tau_{thix}} - \lambda_0\right) e^{-(1+\beta\dot{\gamma}_f \tau_{thix})t/\tau_{thix}} \qquad (3)$$

Where $\lambda_0$ indicates the non-zero (but typically negligibly small) residual value of the thixotropy parameter in the fully-letherized state at $My_i \gg 1$. The expression in eq. (3) is shown in Fig. 7 by the solid lines. The asymptotic value of $\lambda$ at long times depends on the parameter $\beta$ (as evident from the figure) and is given by: $\lambda(\dot{\gamma}_f, t \to \infty) = \frac{1}{1+\beta\dot{\gamma}_f \tau_{thix.}} = \frac{1}{1+\beta My_f}$. The blue dotted line represents the Taylor series expansion series of eq.3 at short times. Solving for the intersection point of the Taylor series expansion at short time and the asymptotic value of $\lambda$ at long time gives the following expression:

$$t^* \cong \frac{\tau_{thix}}{1+\beta My_f}$$

which provides a very easy method for recovering the thixotropic timescale $\tau_{thix}$ (as indicated by the hollow star shown in Fig. 7).

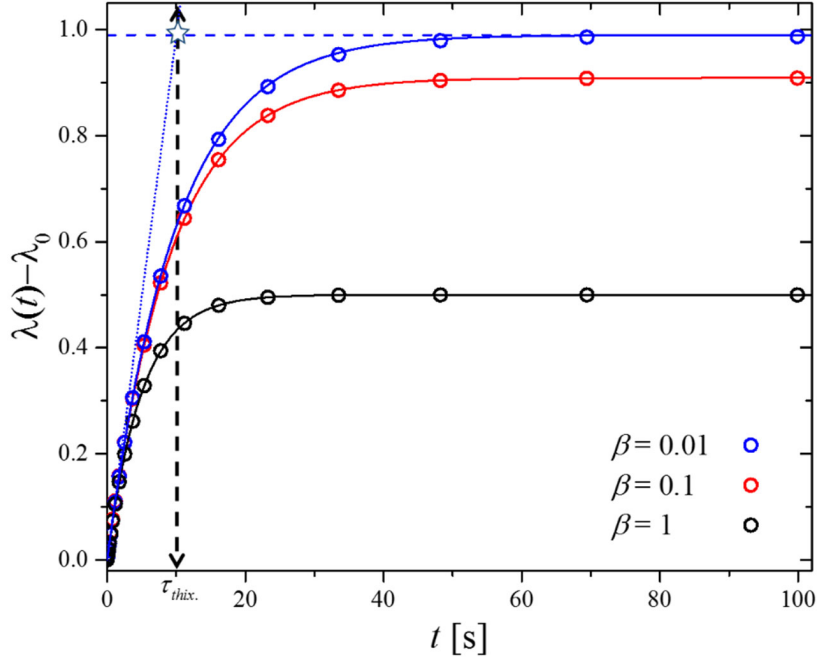

Figure 7. The evolution of the thixotropy parameter, $\lambda$, with total elapsed time of experiment for a very rapid quench/recovery (i.e., a stress jump-down experiment). The hollow circles represent the numerical solution of eq. (2) and the solid line represents the analytical solution given in eq. 3. The hollow star indicates the intersection of the asymptotic value of $\lambda(\dot{\gamma}_f, t \to \infty) = 1/[1+\beta\dot{\gamma}_f \tau_{thix}]$ (blue dashed line) and the Taylor series expansion of eq. (3) (blue dotted line). Provided $My_f \ll 1$ this gives a good direct estimate of the thixotropic timescale, as shown by the black vertical dashed arrow.

## V. Conclusions

In this short article we have proposed a framework and language for quantifying the magnitude of thixotropic effects in complex fluids under steady and time-varying shear flows. We believe the definition and reporting of a *Mnemosyne number*, $My = \tau_{thix}\dot{\gamma}$ (to quantify the relative magnitude of thixotropic effects), as well as a *mutation number* $Mu = t_{exp}/\tau_{thix}$ (to report the duration of a particular test protocol or flow process) will help rheologists unravel and understand the complexities of thixotropic effects in complex fluids. We have outlined several possible rheometric test protocols that enable the thixotropic time scale and the magnitude of the rheological hysteresis in a given material to be quantified in a convenient manner for experimentalists and theoreticians alike. The ramp-down/ramp-up protocol corresponds to a series of up/down vertical trajectories through the *Mu–My* state space sketched in Figure 1. There are, of course, many other putative test protocols that can be proposed to explore this space (for example exponential ramps down/up [27], or a staircase series of constant rates that are stepped down/up [17]) and the optimal protocol may be expected to vary for diverse systems ranging from consumer products to foods comprised of structured pastes or jammed emulsions. However, all such protocols can be represented in the *Mu–My* state space, and the fully 'shear-melted' or destructured material state – in which memories of all previous deformation histories are eliminated – is always clearly represented by the fully *letherized* state located in the upper right of this diagram. Finally, we note that for more complex thixotropic elastoviscoplastic (TEVP) material systems (e.g. many colloidal gels) which show rheological aging, and/or pronounced viscoelastic effects, this two-dimensional state space must also be augmented with additional dimensions (capturing the material age and the viscoelastic relaxation time respectively). Agarwal et al. [11] have recently argued for other shearing protocols (such as step strains imposed after cessation of pre-shearing with different waiting times, $t_w$) that can distinguish between rheological aging and nonlinear viscoelasticity, and it will be interesting to represent the resulting material hysteresis in plots of the type we show in Figs 5 & 6 where the appropriate mutation number now becomes $Mu = t_w / \tau_{thix.}$. Thixotropy will again correspond to non-monotonic measures of hysteresis whereas nonlinear viscoelastic effects should lead to a monotonic decrease in the hysteresis with increasing waiting times between experiments.

**Acknowledgments**

Complex fluids research in the Non-Newtonian Fluids (NNF) group at MIT is supported by a gift from Procter & Gamble.

**Electronic Supplementary Information**

I. Choice of Definition in Hysteresis Area Measure

The magnitude of the hysteresis area computed in a down/up shear rate ramp of the form we propose in the main text can depend sensitively on exactly how the hysteresis is defined and computed (especially because rheologists commonly use log-log representations of the evolving flow curve depicting the evolution of $\sigma(\dot{\gamma})$). However the observation of a distinct local maximum in the hysteresis at an intermediate value of the mutation number is robust in experiments [17, 25], computations [25-27] and the model calculations presented here.

To illustrate this sensitivity to choice of hysteresis area measure we also consider a definition based on logarithmic spacing of the stress <u>and</u> the shear rate;

$$A_\sigma = \int_{\dot{\gamma}_f}^{\dot{\gamma}_i} |\Delta \log \sigma(\dot{\gamma})| d\log \dot{\gamma} \tag{A1}$$

The corresponding plot (un-normalized) has units of [Pa s$^{-1}$], and the location of the maxima in the distribution are shifted towards shorter times as shown for example in Figure SI.1.

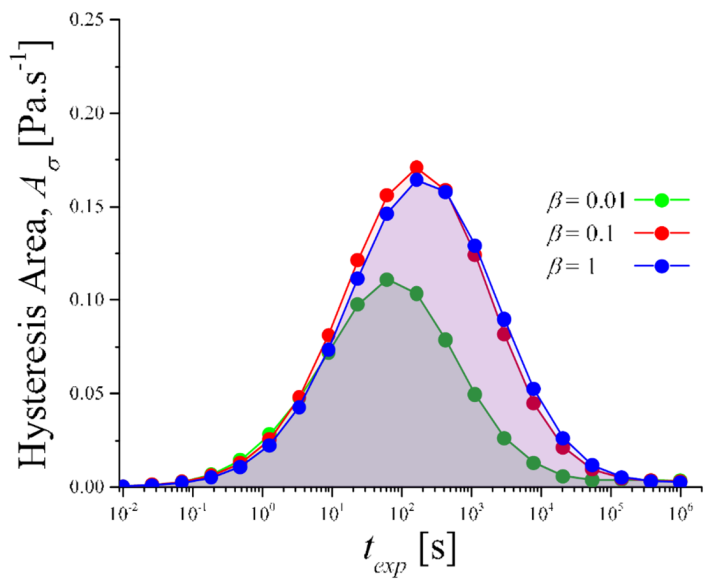

Figure A1. Computed values of the total hysteresis areas measured based on eq. A1 calculated for different experimental times (different mutation numbers), and for different values of parameter $\beta$, for a thixoplastic fluid with similar parameters to those shown in figure 5(c).

Note that the hyteresis areas presented in this figure correspond to a thixoplastic fluid model and thus should be compared to the ones presented in fig. 5c (insert graph) within the main manuscript. It is apparent from this illustration that a clear definition for how a metric such as the hysteresis area is measured or computed during a specified ramp-down/ramp-up protocol will also play an important role in how we determine and report the thixotropic timescale of the material from rheometric experimental protocols.

2. Viscosity Bifurcation Model of Coussot & Bonn

As discussed in the text, many more sophisticated constitutive formulations have been proposed for thixotropic fluids. One example is the 'viscosity bifurcation' model considered by Coussot and Bonn [31]. In this type of model, the time evolution of the microstructure parameter takes a functional form that can be written generically as:

$$\frac{d\xi}{dt} = \frac{1}{\tau_{thix}} - \alpha\xi\dot{\gamma}$$

(A.2)

Note that in this type of model the microstructural parameter for a fully-structured fluid diverges with time when no shear is imposed (i.e., $\xi \to \infty$), and thus we chose to denote this structure parameter as $\xi$, as opposed to the parameter $\lambda$ (used in eq. 3 of the main text) which has a numerical bound of $\lambda = 1$ for a fully structured fluid. Similarly, in eq. (A.2) for clarity in this model we denote $\alpha$ as the dimensionless parameter describing the structure break-up under shear.

The constitutive law for the thixoviscous Coussot-Bonn model can then be written as $\sigma(\xi,t) = \eta_0 f(\xi,\dot{\gamma}(t))\dot{\gamma}(t)$, where $\eta_0$ is the viscosity of a fully destructured fluid ($\xi = 0$). Coussot & Bonn propose a simple diverging function of the form $f(\xi,\dot{\gamma}(t)) = \exp(\xi)$ that captures both the divergence in the viscosity (and possible approach to a yield stress) at low shear rates and also recovers $f \to 1$ when $\xi \to 0$. Other (less strongly) diverging choices of $f$ are also possible in this class of viscosity-bifurcation model (such as power-laws of the form $f = 1 + K\xi^n$).

In Figure A2 we show the hysteretic evolution of the microstructure (fig. A.2a), as well as the resulting shear stress (fig. A.2b) as a function of applied deformation rate for an exponential diverging form of $f(\xi,\dot{\gamma}(t)) = \exp(\xi)$. In contrast to the thixoviscoplastic models considered in the main text, for sufficiently slow ramp rates this viscosity bifurcation model can also predict the evolution of *non-monotonic* flow curves at low shear rates [depending on the value of the $\alpha$ parameter], as indicated by the vertical dotted line in fig. A.2(b). Essentially, for sufficiently slow ramp rates, the rapid growth in the magnitude of structural parameter with time (as shown by the blue curve in fig A.2(a) leads to a divergence in the stress at sufficiently low shear rates.

The details of this non-monotonicity depend sensitively on the functional form of $f(\xi)$ as well as the chosen ramp rate $r$ and the lowest shear rate $\dot{\gamma}_{min}$ specified; however, we again robustly observe a maximum in the hysteresis area when plotted as a function of the experimental duration or mutation number. To illustrate this we show the hysteresis areas computed for our continuous ramp down/up protocols at three different values of $\alpha$ in fig. A.2(c). All of the curves show a distinct local maximum again, as also reported in the main text using other thixoviscous and thixoviscoplastic constitutive equations.

This example further illustrates that the proposed flow protocols, and the expected non-monotonicity in hysteresis area measurements are generic for a wide range of constitutive equations and can be generalized to a broad class of thixotropic fluid models.

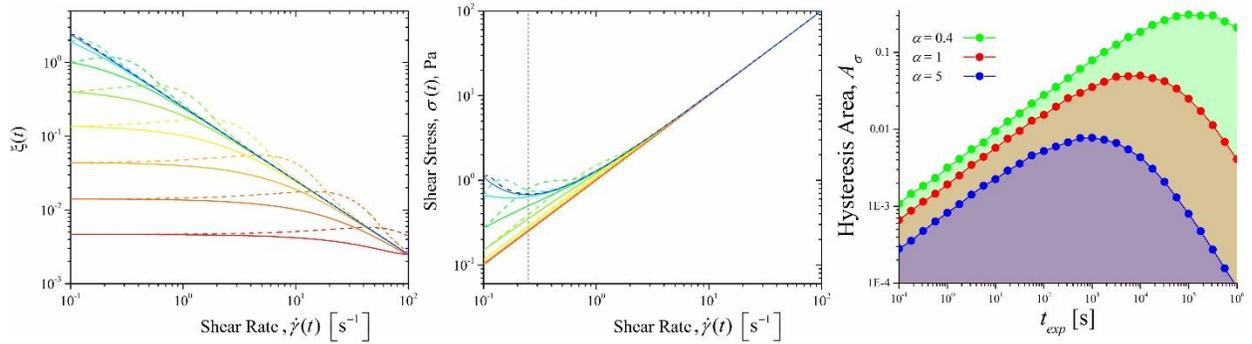

Figure A2. Expected evolution in the microstructural parameter (a) and shear stress (b) vs. applied shear rate, under the ramp-down/ramp-up flow protocol (shown schematically in the top row of Fig.3 of the main paper), for a Coussot-Bonn fluid model with microstructural evolution given by equation A.2. The color increments from blue to red represent longer total times of experiments, $t_{exp}$, corresponding to different *Mu* numbers in the range $Mu = 10^{-2} - 10^5$. (c) The total hysteresis areas $A_\sigma(t_{exp})$ measured based on eq. (2) are calculated for different experimental durations, and different values of the parameter $\alpha$. The solid lines present the initial *ramp-down* flow, and dashed lines

represent the *ramp-up* flow protocols for the same experimental durations. The dotted vertical line in (b) shows the critical shear rate below which a local stress minimum is observed resulting in non-monotonic apparent flow curves.